\documentclass[twocolumn,aps,prl,superscriptaddress]{revtex4}
\usepackage{amsmath}
\usepackage{amssymb}
\usepackage{graphicx}
\makeatletter
\usepackage{epsfig}
\usepackage{color}
\usepackage{ulem}

\makeatother

\begin{document}
\title{A unified theory of thin film and bulk bilayer nickelates}
\author{Jiangfan Wang}
\email[]{jfwang@hznu.edu.cn}
\affiliation{School of Physics, Hangzhou Normal University,  Hangzhou, Zhejiang 311121, China}
\author{Sheng-Yu Yuan}
\affiliation{School of Physics, Hangzhou Normal University,  Hangzhou, Zhejiang 311121, China}
\author{Yi-feng Yang}
\email[]{yifeng@iphy.ac.cn}
\affiliation{Beijing National Laboratory for Condensed Matter Physics and Institute
of Physics, Chinese Academy of Sciences, Beijing 100190, China}
\affiliation{School of Physical Sciences, University of Chinese Academy of Sciences, Beijing 100049, China}

\date{\today}

\begin{abstract}
The discovery of bilayer nickelate superconductivity in both pressurized bulk and thin films has drawn enormous attention on their similarity and distinction. Here we provide a unified explanation based on the two-component scenario for a number of key experimental observations reported recently. Our theory predicts two superconducting domes upon electron or hole doping, separated by a valence bond state near $d_{z^2}$ half filling for strong interlayer superexchange coupling $J$, and a single dome across half filling with a lower $T_c$ for weak or moderate $J$. Increasing doping drives the normal state from a Fermi liquid to non-Fermi liquid or weak insulating behaviors, with quasi-linear-in-$T$ scattering rate near optimal $T_c$, while breaking the interlayer valence bonds by oxygen vacancies or chemical substitution simultaneously suppresses the superconductivity and causes local Kondo scattering of $d_{x^2-y^2}$ electrons. These explain the different superconducting transitions and normal states in bulk and thin films, the effect of $d_{z^2}$ hole or electron doping, and the Kondo effect in non-superconducting samples. We propose bulk superconductivity at ambient pressure by doping or reducing the interlayer magnetic coupling and predict even higher $T_c$ upon electron doping.
\end{abstract}

\maketitle

The discovery of high-$T_c$ superconductivity in the Ruddlesden-Popper (RP) bilayer nickelates {\it R}$_{3}$Ni$_2$O$_7$ in both pressurized bulk \cite{MWang2023Nature,JGCheng2023,HQYuan2024,ZChen2024,JGCheng2024Nature,HHWen2024,XJZhou2024,LShu2024,DLFeng2024,MWang2024,MWang2024b,LYang2024,Shen2025,JJZhang2025,MWang2025,XHChen2025review,KeWang2026,GHCao2026} and compressively strained thin film samples \cite{Hwang2025,Chen2025,Hwang2025b,QKXue2025,ZXShen2025,Hwang2025FL,He2025pseudogap,HHWen2025,ZYChen2025,Nie2025Sr,Nie2025PRL_crossover,HHWen2026pressure,ZYChen60K,Hwang2025Reducing,Nie2025ARPES,HHWen2024Kondo,Kumar2026,Hwang2026halfdome,KuiJin2026,ZYChen2026Three,KJZhou2026} has attracted tremendous interest. Despite  numerous experimental and theoretical investigations \cite{DXYao2023,Dagotto2023,Werner2023,Leonov2023,Eremin2023,YFYang2023,QQin2023,GMZhang2023,TZhou2023,FYang2023,QHWang2023,YHZhang2023,YZYou2023,YYCao2024,CJWu2024PRL,CJWu2024,GSu2024,Bohrdt2024,DXYao2024tJ,ZYWeng2024,ZYLu2024,Kuroki2024,WLi2024,KJiang2024,WKu2024,TXiang2024,QQin2024,WQChen2025,JPHu2025,QHWang2025,JPHu2025PRL,Wang2025,WWu2025,YFYang2025CPL,CJWu2025,GSu2025,DXYao2025film,DXYao2025pressure,JFWangnpjQM26,Watanabe2026}, a consensus regarding the microscopic mechanism of its superconductivity and nontrivial normal state properties has not yet been reached. Key questions remain open: Can the bulk superconduct at ambient pressure? Can thin films have a higher $T_c$ comparable to that of the bulk? Is the $d_{z^2}$ hole pocket ($\gamma$) essential for the superconductivity? To address these issues, a unified theory should be developed under the constraints imposed by latest experimental advances:

1) \textit{Superconductivity. }The bulk shows a transition temperature around $80$ K at high pressures, with a record high $T_c\approx 96$ K achieved recently by chemical substitution to reduce the interlayer distance \cite{JJZhang2025,MWang2025}. Thin films at ambient pressure can be made superconducting but exhibit a much lower $T_c$ of about 40 K \cite{Hwang2025,Chen2025}. The latter may be associated with the elongation of vertical Ni-O-Ni bonds. However, under hydrostatic pressure, the thin film $T_c$ also increases \cite{HHWen2026pressure}. A universal relation has been found that the maximum $T_c$ decreases almost linearly with the $c$-axis lattice constant for both bulk and thin film samples \cite{MWang2025}, indicating positive correlation between $T_c$ and the interlayer superexchange coupling. 

2) \textit{Gap structures. }Scanning tunneling microscopy (STM) measurements of La$_2$PrNi$_2$O$_7$ thin films  support an anisotropic $s^{\pm}$-wave gap symmetry \cite{HHWen2025}. Direct angle-resolved-photoemission spectroscopy (ARPES) measurements on (La,Pr,Sm)$_3$Ni$_2$O$_7$ film suggest a weakly anisotropic $s$-wave gap on the $d_{x^2-y^2}$ anti-bonding band ($\beta$) and an isotropic gap on $\gamma$ \cite{ZYChen2025}. Gap-like features have also been observed on the $d_{x^2-y^2}$ bonding band ($\alpha$) and $\beta$ band in (La,Si)$_3$Ni$_2$O$_7$ films \cite{Nie2025ARPES}. Recent measurements confirm the gap opening on all three Fermi pockets and report a large gap ratio $2\Delta/k_\text{B}T_c\approx 8$ \cite{ZYChen2026Three}, as predicted earlier in the two-component theory \cite{QQin2023}.

3) \textit{Normal states. }While the bulk La$_{3}$Ni$_2$O$_7$ exhibits perfect linear-in-$T$ resistivity above optimal $T_c$ \cite{HQYuan2024,Shen2025}, most thin films show Fermi liquid (FL) normal states \cite{Hwang2025,Chen2025,Hwang2025FL}. Recently, high quality (La,Pr)$_3$Ni$_2$O$_7$ films grown on SrLaAlO$_4$ substrates are found to display quasi-linear-in-$T$ resistivity in samples with the highest $T_c$ \cite{ZYChen60K}. Similar non-Fermi liquid (NFL) behaviors have also been reported in thin films on LaAlO$_3$ substrates  \cite{Hwang2025Reducing,Kumar2026}. In addition, by tuning Sr doping \cite{Nie2025PRL_crossover,Nie2025Sr} or hydrostatic pressure \cite{HHWen2026pressure}, a crossover from metallic to weakly insulating behaviors ($\sim-\ln T$) has been observed. In particular, under pressure tuning, the normal state evolves continuously from FL to NFL as $T_c$ increases to its maximum, and then becomes weakly insulating as $T_c$ drops \cite{HHWen2026pressure}. 

4) \textit{Doping effects. }First-principles calculations of bulk La$_{3}$Ni$_2$O$_7$ suggest the presence of $d_{x^2-y^2}$ bonding ($\alpha$) and anti-bonding ($\beta$) Fermi surfaces and $d_{z^2}$ hole pockets ($\gamma$) at high pressure \cite{MWang2023Nature,DXYao2023,Werner2023}, while ARPES measurement only reported $\alpha$ and $\beta$ Fermi surfaces at ambient pressure \cite{XJZhou2024}, implying the crucial role of $d_{z^2}$ metallization for the superconductivity. By contrast, ARPES measurements on (La,Pr)$_3$Ni$_2$O$_7$ thin films of similar $T_c$ produce conflicting results regarding the existence of $\gamma$ pocket  \cite{QKXue2025,ZXShen2025}, suggesting that the superconductivity of thin films may not be sensitive to the $d_{z^2}$ doping level. Recently, more systematic investigations of thin film superconductors with Sr doping  \cite{Nie2025Sr,Nie2025PRL_crossover}, pressure \cite{HHWen2026pressure}, and oxygen stoichiometry \cite{Nie2025PRL_crossover,Hwang2026halfdome}, all of which may increase the hole doping, consistently reveal a dome-like $T_c$.  

5) \textit{Kondo scattering. }Transport measurements on non-superconducting La$_{3}$Ni$_2$O$_{7-\delta}$ thin films \cite{KuiJin2026,HHWen2024Kondo,Kumar2026} and polycrystalline bulk samples \cite{KeWang2026} reveal logarithmic temperature dependence of the resistivity \cite{KuiJin2026,HHWen2024Kondo,Kumar2026,KeWang2026} as well as negative magnetoresistivity \cite{KuiJin2026,HHWen2024Kondo}, a clear signature of incoherent Kondo effect under oxygen vacancy.

These experimental facts summarize the most essential yet puzzling features of bilayer nickelates, which require consistent explanation within a single theoretical framework. This work provides a unified theory that can satisfactorily address all above observations. Our theory is based on the two-component scenario \cite{YFYang2023,QQin2023,Wang2025}, where the strongly correlated $d_{z^2}$ electrons form interlayer spin-singlet pairing via the Ni-O-Ni superexchange interaction and  hybridize with the more itinerant $d_{x^2-y^2}$ electrons to induce in-plane phase coherence for the superconductivity. A key ingredient to unify the thin film and bulk experiments is to correctly treat the correlation strength of $d_{z^2}$ electrons and hence their metallization through hybridization with the $d_{x^2-y^2}$ bands. 

We propose that all above major physics can be captured by the following two-orbital $t$-$V$-$U$ model: 
\begin{eqnarray}
	H&=&-\sum_{lijs}(t_{ij}+\mu\delta_{ij})c_{lis}^\dagger c_{ljs}-\sum_{lijs}(V_{ij}d_{lis}^\dagger c_{ljs}+H.c.) \notag \\
	&&-t_{\perp}\sum_{is}(d_{1is}^\dagger d_{2is}+h.c.)+U\sum_{li}n_{li\uparrow}^d n_{li\downarrow}^d \label{eq:H},
\end{eqnarray}
where $c_{lis}^\dagger$ ($d_{lis}^\dagger$) creates a $d_{x^2-y^2}$ ($d_{z^2}$) electron of spin $s$ at site $i$ on layer $l$, $n_{lis}^d=d_{lis}^\dagger d_{lis}$ is the $d_{z^2}$ occupation number, $t_\perp$ denotes the interlayer $d_{z^2}$ hopping amplitude, and $V_{i,i+x}=-V_{i,i+y}=V$ is the nearest-neighbor hybridization between two orbitals. The onsite Coulomb repulsion $U$ of the $d_{z^2}$ orbital is included explicitly to tune the correlation strength, while that of $d_{x^2-y^2}$ electrons and the Hund's rule coupling are absorbed in the renormalized tuning parameters in Eq. (\ref{eq:H}) \cite{YYCao2024}. Since the $d_{z^2}$ orbital is nearly half filled and close to the Mott regime, the major effect of the vertical hopping $t_\perp$ is to induce an interlayer superexchange interaction between $d_{z^2}$ spins, namely, $J\sum_{i}\mathbf{S}_{1i}\cdot \mathbf{S}_{2i}$, with $J\propto t_\perp^2/U$. For simplicity, we only keep the nearest-neighbor $t_{ij}=t$ for $d_{x^2-y^2}$ and set it as the energy unit. The chemical potential is fixed such that the $d_{x^2-y^2}$ varies slightly around quarter filling.

\begin{figure}[t]
	\begin{centering}
		\includegraphics[width=0.49\textwidth]{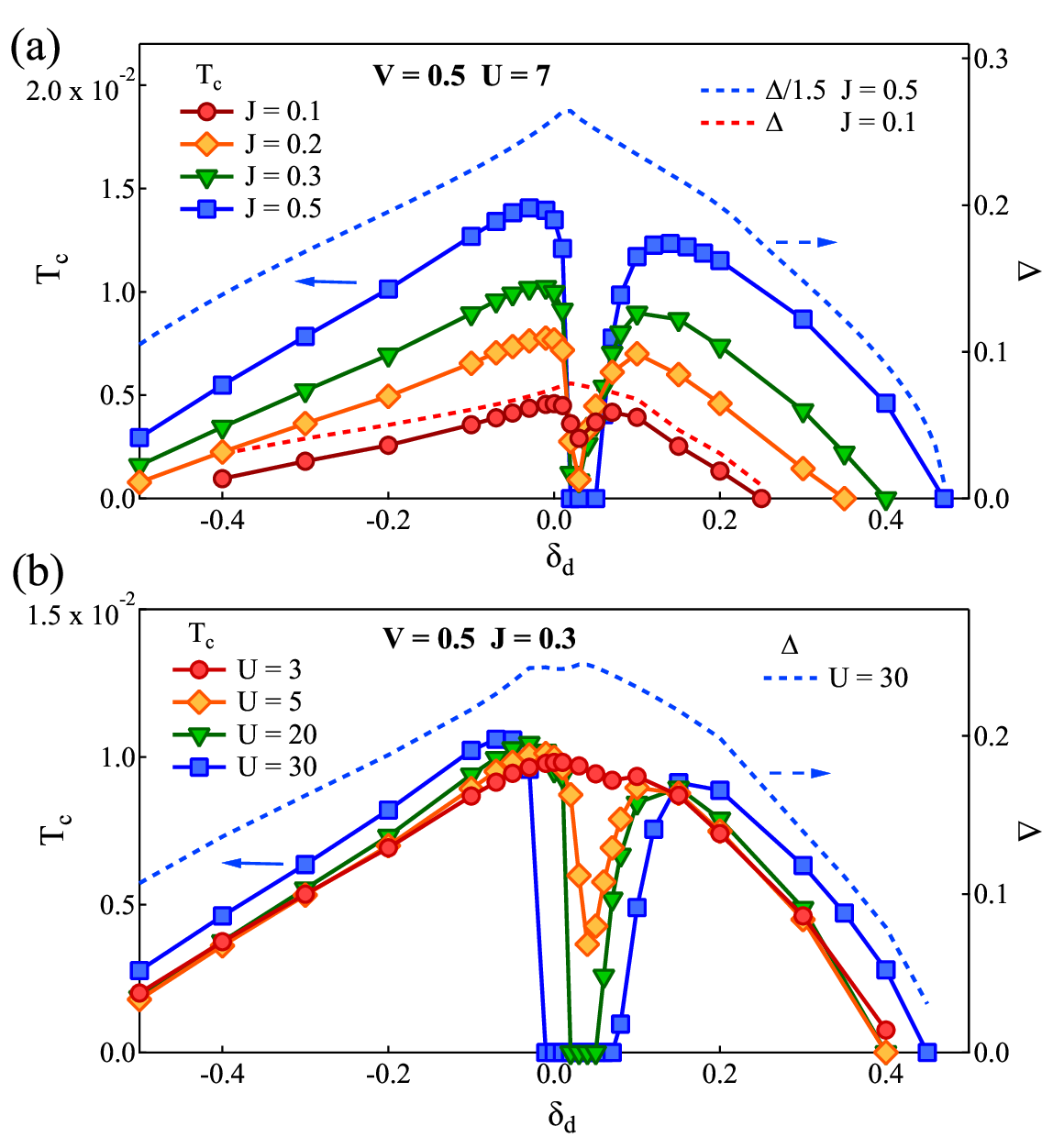}
		\par\end{centering}
	\caption{(a) Superconducting transition temperature $T_c$ as functions of $d_{z^2}$ hole ($\delta_d>0$) or electron ($\delta_d<0$) doping for different $J$ at fixed $V=0.5$, $U=7$. The blue and red dashed lines are the (scaled) spinon valence bond amplitudes $\Delta$ at low temperature $T=0.004$ for $J=0.5$ and $0.1$, respectively. (b) $T_c$ as functions of $\delta_d$ for different $U$ at fixed $V=0.5$, $J=0.3$. The dashed line shows $\Delta$ at $T=0.004$ and $U=30$.}
	\label{fig1}
\end{figure}

\textit{Superconductivity.---}Though simplified, the model still cannot be exactly solved. To treat the magnetic correlation and superconductivity appropriately, we employ the slave particle representation, $d_{lis}=b_{lis}\chi_{li}^\dagger+sb_{li,-s}^\dagger \zeta_{li}$, where $b_{lis}$, $\chi_{li}$ ($\zeta_{li}$) denote the bosonic spinon and fermionic holon (doublon), respectively \cite{Wang2020,Wang2021,Wang2022a,Wang2022b,Wang2025,JFWangnpjQM26,Supp}. $J$ and $U$ are then physical parameters tuning the strength of the interlayer spinon pairs, $\Delta=-\frac{J}{2\mathcal{N}_s}\sum_{is} \langle sb_{1is}b_{2i\bar{s}}\rangle$, and the doublon energy, $U\sum_{li}n_{li}^\zeta$, which, together with the constraint, $n_{li}^\chi+n_{li}^b+n_{li}^\zeta=1$ and $\delta_d=(2\mathcal{N}_s)^{-1}\sum_{li}(\langle n_{li}^\chi\rangle-\langle n_{li}^\zeta\rangle)$, controls the doping level $\delta_d$ and  renormalization of $d_{z^2}$ electrons ($\mathcal{N}_s$ is the number of lattice sites). The self-energies and Green's functions of all particles are obtained self consistently, after which all physical quantities can be calculated \cite{Supp}. The superconducting properties are studied via the interlayer pairing vertex of $d_{x^2-y^2}$ electrons using the Bethe-Salpeter equation, whose momentum dependence contains information of the pairing symmetry. Due to hybridization anisotropy, the pairing vertex takes the form $\Gamma_{ss'}(\mathbf{k},i\omega_n;\mathbf{k}',i\omega_{n'})=ss'\xi_{\bf k}^2\xi_{\mathbf{k}'}^2\tilde{\Gamma}(i\omega_n,i\omega_{n'})$ \cite{Supp}, where the prefactor $\xi_{\bf k}^2=(\cos k_x-\cos k_y)^2$ immediately indicates anisotropic $s^\pm$-wave pairing gaps with opposite signs and nodes along the zone diagonal on $\alpha$ and $\beta$ Fermi surfaces. The nodes are not protected by symmetry and may become minima by higher-order effects such as direct interlayer pairing of $d_{x^2-y^2}$ or disorders, as observed in experiments \cite{HHWen2025,ZYChen2025,ZYChen2026Three}. The interlayer pairing of $d_{z^2}$ electrons is completely local, $\sum_s\langle sd_{1is}d_{2j\bar{s}} \rangle \propto \delta_{ij}$, explaining the isotropic $s$-wave gap on the $\gamma$ pocket \cite{ZYChen2025}. The $t_\perp$ term also induces onsite pairing of higher order ($\propto |\delta_d|^2$, where $\delta_d$ is the $d_{z^2}$ hole/electron doping), which is dropped to simplify the computations. The superconducting transition temperature $T_c$ is determined by the divergence of  $\tilde{\Gamma}(i\omega_n,i\omega_m)$ at the minimal frequency $|\omega_n|=|\omega_m|=\pi T$ and then multiplied by a correction factor of 0.3 to tentatively account for the effect of phase fluctuations based on previous comparisons with Monte Carlo simulations \cite{QQin2023,JFWangnpjQM26,Wang2025}. 

Figure \ref{fig1}(a) plots the resulting $T_c$ as a function of $d_{z^2}$ doping $\delta_d$ for different $J$ and fixed $V=0.5$, $U=7$. For large $J=0.5$ with strong pairing strength, we obtain two separate superconducting domes on electron ($\delta_d<0$) and hole ($\delta_d>0$) doping regions, with a slightly higher maximal $T_c$  on the electron doping side. While the highly asymmetric superconducting dome under hole doping is consistent with the right triangle shape reported in pressurized bulk, the superconductivity under electron doping was only speculated recently in experiments and has yet to be fully explored \cite{Hwang2026halfdome}. The suppression of $T_c$ at large hole/electron doping originates from the reduced spinon pairing amplitude $\Delta$, as demonstrated by the blue dashed line in Fig. \ref{fig1}(a). In real materials, the effective interlayer coupling is also reduced by the factor $(1-|\delta_d|)^2$ \cite{Devereaux2013}, hence $T_c$ is expected to diminish more rapidly upon heavy electron or hole doping. Close to $d_{z^2}$  half filling ($\delta_d=0$), $\Delta$ reaches its maximum while $T_c$ rapidly drops to zero, suggesting an opposite limit where the strong interlayer $d_{z^2}$ singlets are effectively decoupled from $d_{x^2-y^2}$ electrons and form a non-superconducting valence bond state (VBS). Thus unlike cuprates, long-range magnetic orders in bilayer nickelates may only emerge from residual in-plane magnetic interactions. It is easy to estimate that the induced Ruderman-Kittle-Kasuya-Yosida (RKKY) interaction is only a few meV, in agreement with RIXS and neutron measurements of the in-plane magnetic couplings \cite{DLFeng2024,MWang2024}.

As $J$ decreases, the maximal $T_c$ also decreases, implying a positive connection between maximal $T_c$ and $J$ \cite{QQin2023,QQin2025}. However, near half filling, the VBS is also weakened at small $J$ and can now couple with $d_{x^2-y^2}$ electrons to form superconductivity. This leads to a finite $T_c$ around $\delta_d=0$ that merges the two superconducting domes into a single one. For $J=0.1$ and large $U$, there is a local minimum of $T_c$ near half filling, which disappears completely for smaller $J$ or moderate $U$. $T_c$ now follows the valence bond amplitude $\Delta$ more closely, as shown by the red dashed line in Fig. \ref{fig1}(a). The smaller but finite $T_c$ that is less sensitive to $d_{z^2}$ hole doping agrees well with recent experimental observations in thin films \cite{QKXue2025,ZXShen2025}, where first-principles calculations indeed predict a smaller superexchange interaction $J$ due to the elongated vertical Ni-O-Ni bonds \cite{DXYao2025film}. The nonmonotonic evolution of $T_c$ with hole doping at $J=0.1$ is consistent with the observed dome-like shape in thin films upon Sr doping \cite{Nie2025Sr}, hydrostatic pressure \cite{HHWen2026pressure}, and increasing oxygen stoichiometry \cite{Hwang2026halfdome}. 

Figure \ref{fig1}(b) plots the calculated $T_c$ as functions of $\delta_d$ for different $U$ at fixed $V=0.5$ and $J=0.3$. Similarly, one finds two separated domes at large $U$ and a single one at moderate $U$. The opposite trends of $\Delta$ and $T_c$ as $\delta_d$ approaches zero again suggests the formation of VBS at large $U$ where $d_{z^2}$ electrons are well localized. Therefore, tuning $J$ and $U$ can drive a phase transition between the VBS and superconductivity at fixed $d_{z^2}$ occupancy around half filling, which is in some sense similar to the local-to-itinerant transition in heavy fermion systems. Their opposite influence on the superconductivity at large and small $|\delta_d|$ is closely associated with the delocalization of $d_{z^2}$ orbitals. When the $d_{z^2}$ electrons are well delocalized at large $|\delta_d|$, they promote the interlayer pairing and hence the maximal $T_c$; but when $|\delta_d|$ is small, the superconductivity can only emerge at moderate $U$ and $J$ where the $d_{z^2}$ quasiparticles can already exist and hybridize with $d_{x^2-y^2}$. This provides a unified explanation of the bulk and thin film experiments, where the thin films can become superconducting even at ambient pressure but has a lower maximal $T_c$.

\begin{figure}[t]
	\begin{centering}
		\includegraphics[width=0.51\textwidth]{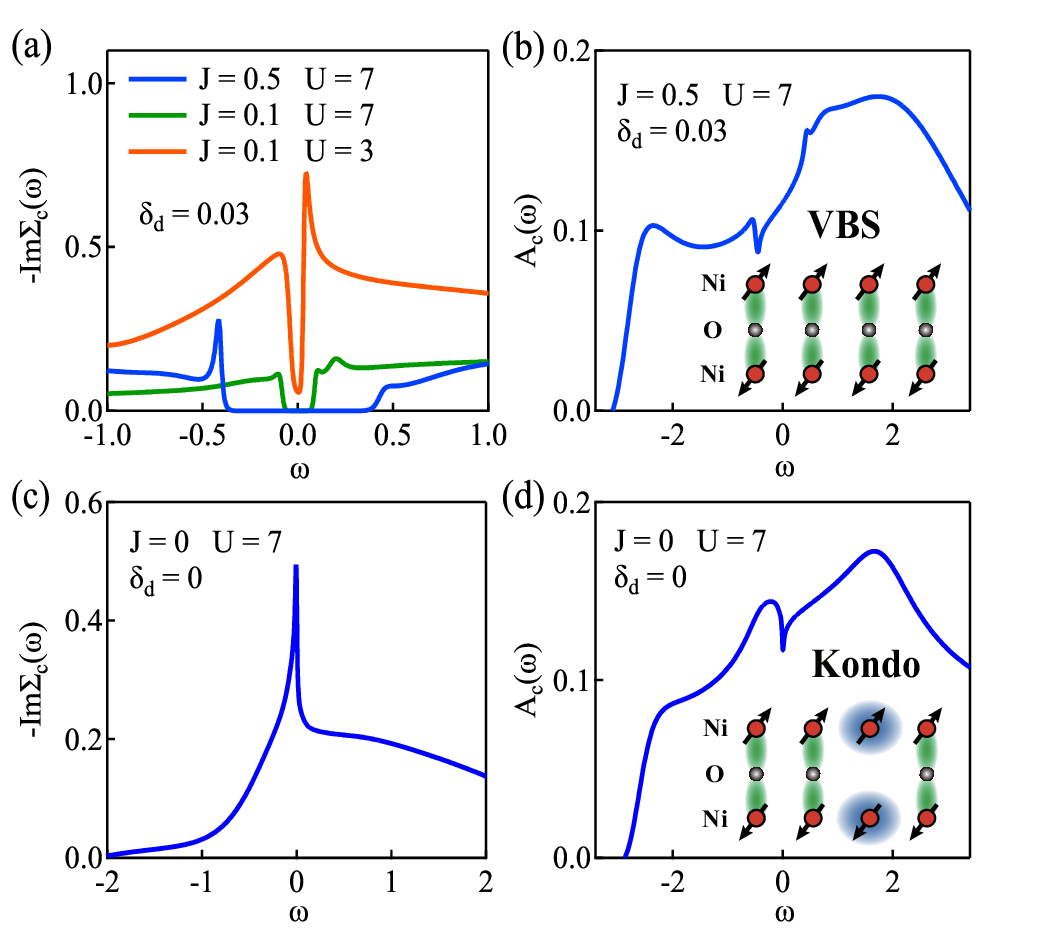}
		\par\end{centering}
	\caption{(a) The imaginary part of the $d_{x^2-y^2}$ self-energy at $\delta_d=0.03$, showing the VBS gap evolution for different $J$ and $U$. (b) The $d_{x^2-y^2}$ density of states $A_c(\omega)$ at $J=0.5$ and $U=7$. (c)(d) The $d_{x^2-y^2}$ self-energy and density of states at $J=0$, $U=7$ for $\delta_d=0$, showing the occurrence of Kondo effects. Other parameters are fixed as $T=0.004$,  $V=0.5$. The insets of (b) and (d) illustrate the vertical Ni-O-Ni bonds in stoichiometric system and with inner apical oxygen vacancies, respectively.}
	\label{fig2}
\end{figure}

\textit{VBS and Kondo effect.---}The VBS has important influences on $d_{x^2-y^2}$ electrons, as may be seen from their self-energy, $\Sigma_c(\mathbf{k},\omega)=\xi_{\mathbf{k}}^2\Sigma_c(\omega)$. As shown in Fig. \ref{fig2}(a) for $J=0.5$ and $U=7$ at $\delta_d=0.03$, there is a VBS gap around zero frequency in $-\text{Im}\Sigma_c(0)$, indicating effective decoupling of two orbitals and hence absence of electron scattering at low temperature. For small interlayer coupling ($J$) or  correlation strength ($U$), the interlayer valence bonds can be easily broken and the gap turns into a dip, as shown in Fig. \ref{fig2}(a) for $J=0.1$ and $U=3$. The hybridization between $d_{x^2-y^2}$ and $d_{z^2}$ quasiparticles then leads to superconductivity even at half filling.

When there exist inner apical oxygen vacancies or one of the two Ni atoms is replaced by a nonmagnetic atom such as aluminium \cite{GHCao2026}, the VBS is destroyed to produce one or two decoupled Ni-$d_{z^2}$ spins, whose hybridization with surrounding $d_{x^2-y^2}$ electrons can induce effective Kondo scattering in bilayer nickelates (see inset of Fig. \ref{fig2}(d)). One may then expect characteristic  $-\ln T$ resistivity and negative magnetoresistivity. Since the breakdown of the VBS simultaneously destroys the interlayer pairing, it also disfavors the superconductivity, thus explaining the observed Kondo physics in non-superconducting thin films and their seeming competition \cite{HHWen2024Kondo,KuiJin2026,KeWang2026,Kumar2026}. Numerically, one can simulate this by setting the local $J$ to zero. Figures \ref{fig2}(c) and \ref{fig2}(d) show the imaginary part of the $d_{x^2-y^2}$ self-energy and the density of states (DOS), respectively, for $J=\delta_d=0$, $V=0.5$ and $U=7$ at $T=0.004$. We see a sharp peak in $\text{Im}\Sigma_c(\omega)$ and a dip in the DOS at zero frequency, a clear indication of Kondo resonance in contrast to the VBS gap at large $J$.

\begin{figure}[t]
	\begin{centering}
		\includegraphics[width=0.48\textwidth]{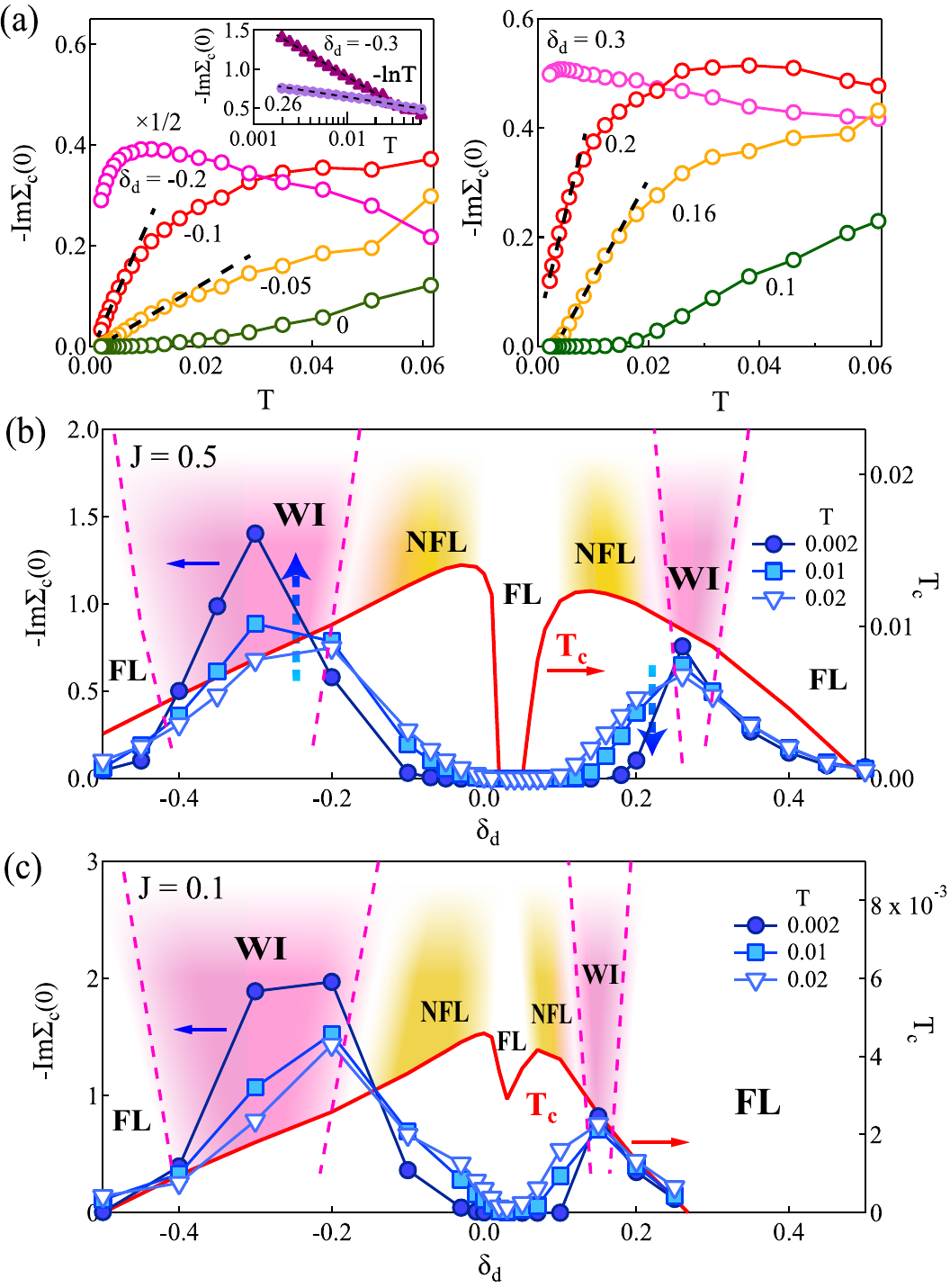}
		\par\end{centering}
	\caption{(a) Imaginary part of $d_{x^2-y^2}$ self-energy at zero frequency, $-\text{Im}\Sigma_c(0)$, as functions of the temperature for different electron (left panel) or hole (right panel) doping $\delta_d$ at $J=0.5$, $V=0.5$, $U=7$. The inset shows the $-\ln T$ dependence at $\delta_d=0.26$ and -0.3. (b)(c) Comparison of $T_c$ and $-\text{Im}\Sigma_c(0)$ as functions of $\delta_d$ at different low temperatures for $J=0.5$, 0.1 at $V=0.5$, $U=7$. The different colors represent the Fermi liquid (FL), non-Fermi liquid (NFL), and weakly insulating (WI) normal states. The dashed lines mark the crossover between different regions.}
	\label{fig3}
\end{figure}

\textit{Normal state.---}The normal state properties of stoichiometric compound can be studied by calculating the $d_{x^2-y^2}$ self-energy at zero frequency from the normal state solution, which represents the quasiparticle scattering rate due to electron correlations. Figure \ref{fig3}(a) shows the temperature dependence of $-\text{Im}\Sigma_c(0)$ on  electron (left panel) and hole (right panel) doping sides at $J=V=0.5$ and $U=7$. Close to half filling, it vanishes at low temperatures due to the VBS gap, indicating a Fermi liquid normal state. Increasing hole (electron) doping quickly suppresses the VBS gap and enhances the inter-orbital scattering, causing a NFL normal state at low temperature. Comparison with the $T_c$ curves in Fig. \ref{fig3}(b) finds a  quasi-linear-in-$T$ behavior  around the optimal doping \cite{Wang2025}. As $|\delta_d|$ further increases, a weakly insulating (WI) region emerges where $-\text{Im}\Sigma_c(0)$ increases logarithmically with decreasing temperature, as shown in the inset of Fig. \ref{fig3}(a) for $\delta_d=-0.3$ and $0.26$. The temperature where it reaches the maximum defines a characteristic scale, above which the interlayer valence bonds are destroyed by thermal fluctuations (not oxygen vacancies) to cause incoherent Kondo scattering of conduction electrons. Consequently, the $d_{x^2-y^2}$ self-energy exhibits a small peak at zero frequency, and the quasiparticle spectra disappears along the $k_x$ and $k_y$ axes, marking potential Fermi surface reconstruction across the WI region \cite{Supp}. Indeed, as $|\delta_d|$ further increases, the system becomes a hybridized Fermi liquid, where the $d_{z^2}$ electrons form well defined quasiparticle bands hybridizing with the $d_{x^2-y^2}$ electrons. These phenomena accompanying with the overall doping-induced delocalization process of $d_{z^2}$ are insensitive to the $d_{x^2-y^2}$ filling \cite{Supp}, reflecting their robustness under varying experimental conditions. As shown in Fig. \ref{fig3}(c), they also persist for small $J=0.1$, except that their boundaries on the hole doping side shift towards smaller $\delta_d$, indicating that the $d_{z^2}$ VBS are much weaker and can be delocalized by smaller hole doping. Experimentally, the evolution from FL to NFL normal states as $T_c$ increases has been observed in thin films by varying oxygen content \cite{ZYChen60K} or increasing pressure \cite{HHWen2026pressure}. A crossover from metallic to WI behavior with $\ln T$ resistivity as the hole doping increases is also reported in Sr doped \cite{Nie2025PRL_crossover,Nie2025Sr} and pressurized thin films \cite{HHWen2026pressure}. The WI region, however, has not been observed in bulk, possibly due to its larger $J$ and the limited range of $d_{z^2}$ hole density by pressure tuning \cite{QHWang2025,DXYao2025pressure}. It is important to clarify the origin of the WI behavior by excluding potential disorder effects in future experiment.

\begin{figure}[t]
	\begin{centering}
		\includegraphics[width=0.49\textwidth]{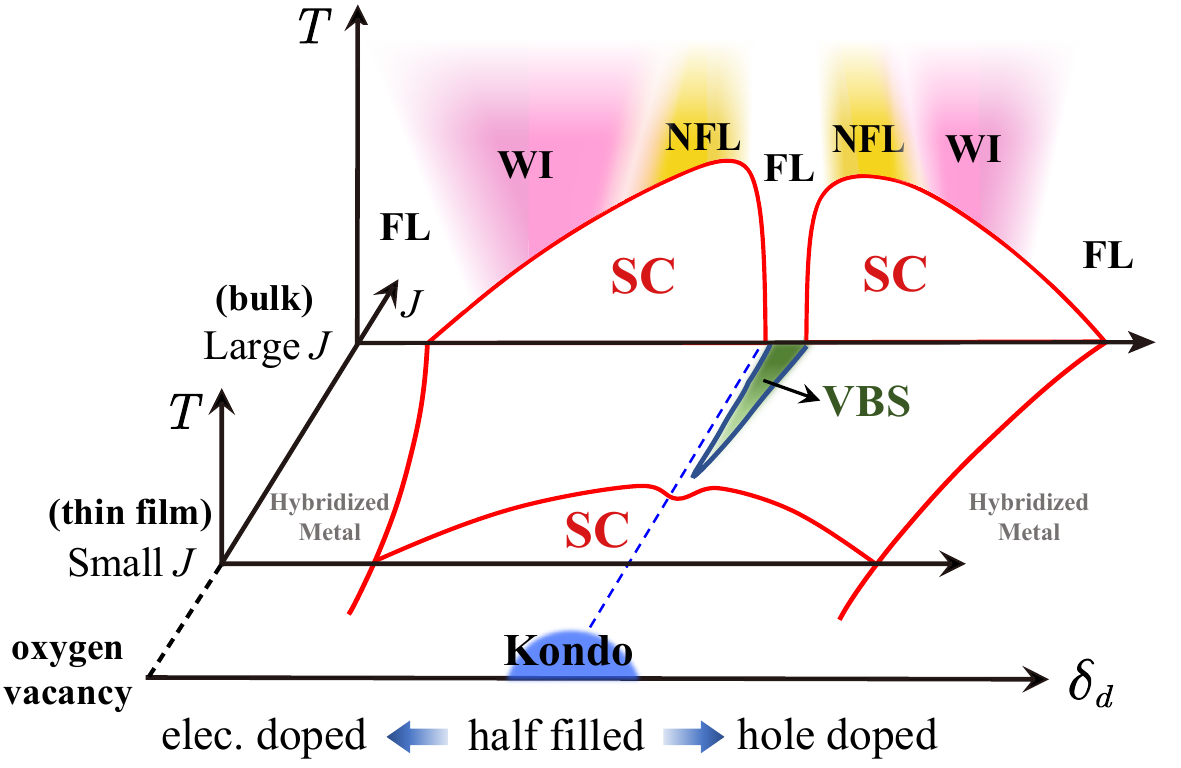}
		\par\end{centering}
	\caption{A unified theoretical phase diagram of bilayer nickelates marking different phases in terms of the superexchange $J$, $d_{z^2}$ hole/electron doping $\delta_d$, and the temperature $T$. The orange and pink colors in the normal state correspond to the quasi-linear-in-$T$ and weakly insulating behaviors of $-\text{Im}\Sigma_c(0)$ in the normal state, respectively. The large and small $J$ correspond to the bulk and compressively strained thin film bilayer nickelates, respectively.}
	\label{fig4}
\end{figure}

\textit{Global phase diagram.---}We can now construct a unified phase diagram for bulk and thin film bilayer nickelates in terms of the interlayer superexchange coupling, $d_{z^2}$ doping, and the temperature, as plotted schematically in Fig. \ref{fig4}. For large $J$ (bulk), there exist two superconducting domes separated by an intermediate VBS near half filling, where the weakly interacting $d_{x^2-y^2}$ electrons effectively decouple from the $d_{z^2}$ VBS and cause a Fermi liquid normal state or a magnetic state due to residual in-plane magnetic couplings such as the RKKY interaction. Increasing electron or hole doping enhances the hybridization and leads to the superconductivity. The normal state shows NFL behaviors around the optimal $T_c$. The WI region at larger doping may not be reached due to the limited range of $d_{z^2}$ hole density upon pressure tuning, leading to a wide range of NFL above $T_c$ in pressurized bulk. For small $J$ (thin film), the $d_{z^2}$ valence bonds are weakened, and the hybridization can lead to superconductivity even at half filling. The WI region shifts to smaller doping and explains the experimental observation in Sr-doped and pressurized thin films. Further increasing doping suppresses the superconductivity and turns the system into a hybridized metal. Inner apical oxygen vacancies or replacing Ni by non-magnetic atoms destroys the interlayer VBS, which not only breaks the superconducting pairing but also produces local Ni spins that explains the Kondo effects in non-superconducting films. All these seemingly unrelated phenomena are now understood within a single unified framework.

Our theory has two immediate predictions: 1) ambient pressure superconductivity by doping or reducing the interlayer magnetic coupling through, for example, stretching along $c$ axis or chemical substitution to increase the interlayer distance; 2) a second superconducting dome upon electron doping with an even higher maximal $T_c$.  Previously, we have predicted $T_c^{\rm max}\approx 0.05 J$ under optimized conditions \cite{QQin2023,JFWangnpjQM26}, in good agreement with bulk experiments. Near half filling, Fig. \ref{fig1} predicts an opposite trend for large to moderate $J$ and hence a maximum $T_c$ of about 0.01 at moderate $U$, which corresponds to about 50 K for realistic hopping parameter $t$. The current thin film superconductivity at ambient pressure seems already close to this maximum $T_c$, and only joint tuning of $d_{z^2}$ doping and interlayer distance or in-plane parameters may raise it to the bulk level.

This work was supported by the National Natural Science Foundation of China (Grants No. 12474136 and No. 12304174).

\end{document}


\title{A unified theory of thin film and bulk bilayer nickelates\\
\vspace{0.2cm}
- Supplemental Material -}
\author{Jiangfan Wang}
\affiliation{School of Physics, Hangzhou Normal University,  Hangzhou, Zhejiang 311121, China}
\author{Sheng-Yu Yuan}
\affiliation{School of Physics, Hangzhou Normal University,  Hangzhou, Zhejiang 311121, China}
\author{Yi-feng Yang}
\affiliation{Beijing National Laboratory for Condensed Matter Physics and Institute
	of Physics, Chinese Academy of Sciences, Beijing 100190, China}
\affiliation{School of Physical Sciences, University of Chinese Academy of Sciences, Beijing 100049, China}

\maketitle

\subsection{I. Dynamic Schwinger Boson Approach}\label{sec:SB}

In our approach, the strongly correlated $d_{z^2}$ electron operator is represented as 
\begin{equation}
	d_{lis}=\chi_{li}^\dagger b_{lis}+sb_{li,-s}^\dagger \zeta_{li}
\end{equation}
where $b_{lis}^\dagger$ creates bosonic spinons, and $\chi_{li}^\dagger$ and $\zeta_{li}^\dagger$ create fermionic holons and doublons, respectively. These slave particles satisfy the number constraint $n_{li}^\chi+ n_{li}^b+n_{li}^\zeta\equiv\chi_{li}^\dagger \chi_{li}+\sum_s b_{lis}^\dagger b_{lis}+\zeta_{li}^\dagger \zeta_{li}=1$, which is implemented on average by the following Lagrange multiplier term 
\begin{equation}
	H_\lambda=\lambda\sum_{li}\left(n_{li}^\chi+ n_{li}^b+n_{li}^\zeta-1\right).
\end{equation}
The $d_{z^2}$ number density can be written as $n_{li}^d=\sum_s d_{lis}^\dagger d_{lis}= n_{li}^b+2n_{li}^\zeta$. For fixed $d_{z^2}$ hole concentration $\delta_{d}$, we have $(2\mathcal{N}_s)^{-1}\sum_{li}(\langle n_{li}^b\rangle+2\langle n_{li}^\zeta\rangle )=1-\delta_d$, which leads to $(2\mathcal{N}_s)^{-1}\sum_{li}(\langle n_{li}^\chi\rangle-\langle n_{li}^\zeta\rangle)=\delta_d$ due to the constraint. This is implemented by a chemical potential term
\begin{equation}
	H_{\mu'}=\mu'\sum_{li}\left(n_{li}^\chi-n_{li}^\zeta-\delta_d\right).
\end{equation}
In this representation, the Coulomb interaction becomes the onsite energy of doublon,
\begin{equation}
	H_U=U\sum_{li}n_{li\uparrow}^d n_{li\downarrow}^d=U\sum_{li}\zeta_{li}^\dagger \zeta_{li}.
\end{equation}
The nearest neighbor hybridization between $d_{z^2}$ and $d_{x^2-y^2}$ orbital is expressed as 
\begin{equation}
	H_\text{hyb}=-\frac{2V}{\sqrt{\mathcal{N}_s}}\sum_{ls\mathbf{pq}}\left(b_{l,\mathbf{p+q},s}^\dagger \chi_{l\mathbf{p}}c_{l\mathbf{q}s}+s\zeta_{l,\mathbf{p+q}}^\dagger b_{l\mathbf{p},-s}c_{l\mathbf{q}s}\right)\xi_{\bf q}+H.c.
\end{equation}  
where $\xi_\mathbf{q}=\cos q_x-\cos q_y$. Finally, the spin density of $d_{z^2}$ orbital is written as $\mathbf{S}_{li}=\frac{1}{2}\sum_{ss'}b_{lis}^\dagger \boldsymbol{\sigma}_{ss'}b_{lis'}$, so that the interlayer superexchange can be decomposed as 
\begin{equation}
	J\mathbf{S}_{1i}\cdot \mathbf{S}_{2i}\rightarrow \Delta\sum_{s}sb_{1is}^\dagger b_{2i,-s}^\dagger +H.c.+\frac{2|\Delta|^2}{J},
\end{equation} 
where $\Delta=-\frac{J}{2}\sum_{s}\langle sb_{1is}b_{2i,-s}\rangle$ is the mean-field spinon valence bond amplitude.

Combining all above terms and the kinetic energy of $d_{x^2-y^2}$ orbital gives the following action:
\begin{eqnarray}
	S&=&-\sum_{lks}\bar{c}_{lks}(i\omega_n-\epsilon_{\bf k})c_{lks}-\sum_{lk}\bar{\chi}_{lk}(i\omega_n-\lambda-\mu')\chi_{lk}-\sum_{lk}\bar{\zeta}_{lk}(i\omega_n-\lambda+\mu'-U)\zeta_{lk} \notag \\
	&&-\sum_{lks}\bar{b}_{lks}(i\nu_n-\lambda)b_{lks}+\sum_{ks}\left( s\bar{b}_{1ks}\bar{b}_{2,-k,-s}\Delta+c.c.\right)+2\beta\mathcal{N}_s \left(\frac{|\Delta|^2}{J}-\lambda-\mu'\delta_d\right)\notag \\
	&&-\frac{2V}{\sqrt{\beta\mathcal{N}_s}}\sum_{lspq}\left(\bar{b}_{l,p+q,s} \chi_{lp}c_{lqs}+s\bar{\zeta}_{l,p+q} b_{lp,-s}c_{lqs}\right)\xi_{\bf q}+c.c. \label{eq:S}
\end{eqnarray} 
where we have used the simplified notation $k\equiv(\mathbf{k},\omega_n(\nu_n))$ for fermion (boson), and $\epsilon_{\bf k}=-2t(\cos k_x+\cos k_y)-\mu$ is the dispersion relation of free $d_{x^2-y^2}$ electrons. We fix $\mu=-1.44$ so that the $d_{x^2-y^2}$ orbital is nearly quarter filled. The partition function of the system is then $\mathcal{Z}=\int \mathcal{D}[c,b,\chi,\zeta]\exp\{-S\}$. 

\textit{Self-consistent equations.---}After obtaining the action Eq. ({\ref{eq:S}}), we can derive the self-consistent equations for the self-energies and Green's functions using the Luttinger-Ward functional or Dyson-Schwinger equations. The Green's functions are:
\begin{eqnarray}
	G_c(\mathbf{k},i\omega_n)&=&\frac{1}{i\omega_n-\epsilon_{\bf k}-\xi_{\bf k}^2\Sigma_c(i\omega_n)},\notag \\
	G_\chi(i\omega_n)&=&\frac{1}{i\omega_n-\lambda-\mu'-\Sigma_\chi(i\omega_n)},\notag \\
	G_\zeta(i\omega_n)&=&\frac{1}{i\omega_n-\lambda+\mu'-U-\Sigma_\zeta(i\omega_n)}, \notag \\
	G_b(i\nu_n)&=&\frac{\gamma_b(-i\nu_n)}{\gamma_b(i\nu_n)\gamma_b(-i\nu_n)-|\Delta|^2}, \label{eq:G}
\end{eqnarray}
where $\gamma_b(i\nu_n)=i\nu_n-\lambda-\Sigma_b(i\nu_n)$. The self-energies depend self-consistently on the Green's functions:
\begin{eqnarray}
	\Sigma_c(i\omega_n)&=&\frac{V^2}{\beta}\sum_m\left[G_\chi(i\nu_m-i\omega_n)-G_\zeta(i\nu_m+i\omega_n)\right]G_b(i\nu_m), \notag \\
	\Sigma_\chi(i\omega_n)&=&\frac{2V^2}{\beta}\sum_m \bar{G}_c(i\omega_m)G_b(i\omega_n+i\omega_m),\notag \\
	\Sigma_\zeta(i\omega_n)&=&-\frac{2V^2}{\beta}\sum_m\bar{G}_c(i\omega_m)G_b(i\omega_n-i\omega_m), \notag \\
	\Sigma_b(i\nu_n)&=&-\frac{V^2}{\beta}\sum_m\bar{G}_c(i\omega_m)\left[G_\chi(i\nu_n-i\omega_m)-G_\zeta(i\nu_n+i\omega_m)\right], \label{eq:Sigma}
\end{eqnarray}
where $\bar{G}_c(i\omega_m)=\mathcal{N}_s^{-1}\sum_{\bf k}\xi_{\bf k}^2G_c(\mathbf{k},i\omega_m)$. 

The mean field variables $\lambda$, $\Delta$ and $\mu'$ are determined by minimizing the free energy, leading to three equations:
\begin{eqnarray}
	1&=&\frac{1}{\beta}\sum_n\left(G_\chi(i\omega_n)+G_\zeta(i\omega_n)-2G_b(i\nu_n)\right), \notag \\
	\delta_d&=&\frac{1}{\beta}\sum_n\left(G_\chi(i\omega_n)-G_\zeta(i\omega_n)\right), \notag \\
	\frac{1}{J}&=&\frac{1}{\beta}\sum_n\frac{1}{\gamma_b(i\nu_n)\gamma_b(-i\nu_n)-|\Delta|^2}. \label{eq:con}
\end{eqnarray}  
Equations (\ref{eq:G})-(\ref{eq:con}) are solved self-consistently, after which all physical quantities can be obtained.

\subsection{II. Cooper Instability}\label{sec:CI}

\begin{figure}
	\centering\includegraphics[scale=0.6]{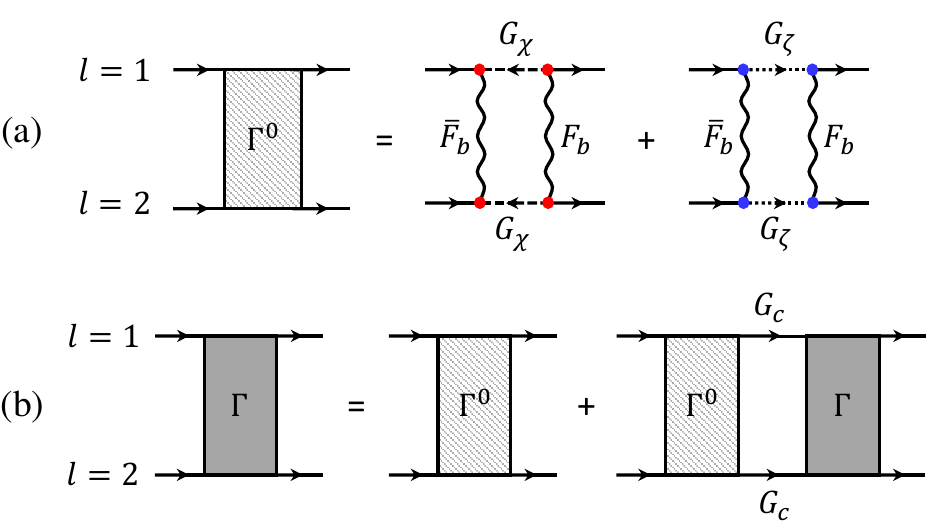}
	\caption{(a) The ``bare" interlayer pairing vertex for $d_{x^2-y^2}$ electrons. (b) The Bethe-Salpeter equation for the full pairing vertex.}
	\label{fig:S1}
\end{figure}

The interlayer pairing vertex of $d_{x^2-y^2}$ electrons can be calculated using the Bethe-Salpeter (BS) equation, 
\begin{eqnarray}
	\Gamma_{ss'}(k,k',q)=\Gamma_{ss'}^0(k,k',q)+\frac{1}{\beta\mathcal{N}_s}\sum_{k''s''}\Gamma_{ss''}^0(k,-k''-q,q)G_c(k''+q)G_c(-k'')\Gamma_{s''s'}(k'',k',q), \label{eq:BS}
\end{eqnarray}
where we have denoted $k\equiv(\mathbf{k},i\omega_n)$ and $q\equiv (\mathbf{q},i\nu_l)$. The ``bare" pairing vertex has the form
\begin{equation}
	\Gamma_{ss'}^0(k,k',q)=ss'\tilde{\Gamma}^0(i\omega_n,i\omega_{n'},i\nu_l)\xi_\mathbf{k+q}\xi_\mathbf{-k}\xi_\mathbf{k'+q}\xi_\mathbf{-k'}, 
	\label{eq:Gmm0}
\end{equation}
where
\begin{eqnarray}
	\tilde{\Gamma}^0(i\omega_n,i\omega_{n'},i\nu_l)&=&\frac{V^4}{\beta}\sum_m F_b(i\omega_m-i\omega_n)\bar{F}_b(i\omega_m+i\omega_{n'}+i\nu_l) \notag\\
	&&\times\left[G_\chi(i\omega_m)G_\chi(-i\omega_m-i\nu_l)+G_\zeta(-i\omega_m)G_\zeta(i\omega_m+i\nu_l)\right]. \label{eq:Gamma0}
\end{eqnarray}
Eq. (\ref{eq:Gamma0}) contains the anomalous Green's function of spinon, $F_b(i\nu_n)=\Delta/(\gamma_b(i\nu_n)\gamma_b(-i\nu_n)-|\Delta|^2)$, and its complex conjugate $\bar{F}_b(i\nu_n)$. Equations (\ref{eq:BS})-(\ref{eq:Gamma0}) can be schematically represented by the Feynman diagrams in Fig. \ref{fig:S1}.

Substituting Eq. (\ref{eq:Gmm0}) into Eq. (\ref{eq:BS}) and assuming static uniform solution $q=(\mathbf{q},i\nu_l)=0$, one has
\begin{equation}
	\Gamma_{ss'}(k,k')=ss'\tilde{\Gamma}(i\omega_n,i\omega_{n'})\xi_{\bf k}^2\xi_{\bf k'}^2
\end{equation}
and 
\begin{eqnarray}
	\tilde{\Gamma}(i\omega_n,i\omega_{n'})=\tilde{\Gamma}^0(i\omega_n,i\omega_{n'})&+&\frac{2}{\beta\mathcal{N}_s}\sum_{\mathbf{k}''n''}\tilde{\Gamma}^0(i\omega_n,-i\omega_{n''})\xi_{\bf -k''}^2\xi_{\bf k''}^2 G_c(\mathbf{k}'',i\omega_{n''}) \notag \\
	&&\times G_c(-\mathbf{k}'',-i\omega_{n''})\tilde{\Gamma}(i\omega_{n''},i\omega_{n'}),
\end{eqnarray}
which can be numerically inversed and yields $\tilde{\Gamma}(i\omega_n,i\omega_{n'})$. The real part of the pairing vertex displays a peak at the minimal Matsubara frequency $|\omega_n|=|\omega_{n'}|=\pi T$. If this peak diverges at low temperature, the system acquires Cooper instability and enters the superconducting state. The momentum-dependent factor $\xi_{\bf k}^2\xi_{\bf k'}^2$ implies an anisotropic $s^{\pm}$-wave gap symmetry on the $\alpha$ and $\beta$ bands. By contrast, the interlayer pairing of $d_{z^2}$ electrons is completely local:
\begin{eqnarray}
	\Delta_{z}(\mathbf{r}_i-\mathbf{r}_j)&\equiv& \sum_s\langle sd_{1is}d_{2j,-s}\rangle=\frac{1}{\mathcal{N}_s}\sum_{s\mathbf{k}}\langle sd_{1\mathbf{k}s}d_{2,\mathbf{-k},-s}\rangle e^{i\mathbf{k}\cdot (\mathbf{r}_i-\mathbf{r}_j)} \notag \\
	&=&\frac{1}{\mathcal{N}_s}\sum_{\mathbf{k}}\Delta_z(\mathbf{k}) e^{i\mathbf{k}\cdot (\mathbf{r}_i-\mathbf{r}_j)}\propto \delta_{ij}.
\end{eqnarray}
Consequently, $\Delta_z(\mathbf{k})=\sum_s\langle sd_{1\mathbf{k}s}d_{2,\mathbf{-k},-s}\rangle=\Delta_z$ is momentum independent, indicating an isotropic $s$-wave gap on the $\gamma$ pockets from the $d_{z^2}$ bonding band, whose superconducting gap has the same sign as that on the $d_{x^2-y^2}$ bonding band $\alpha$.

\begin{figure}[t]
	\centering\includegraphics[scale=0.55]{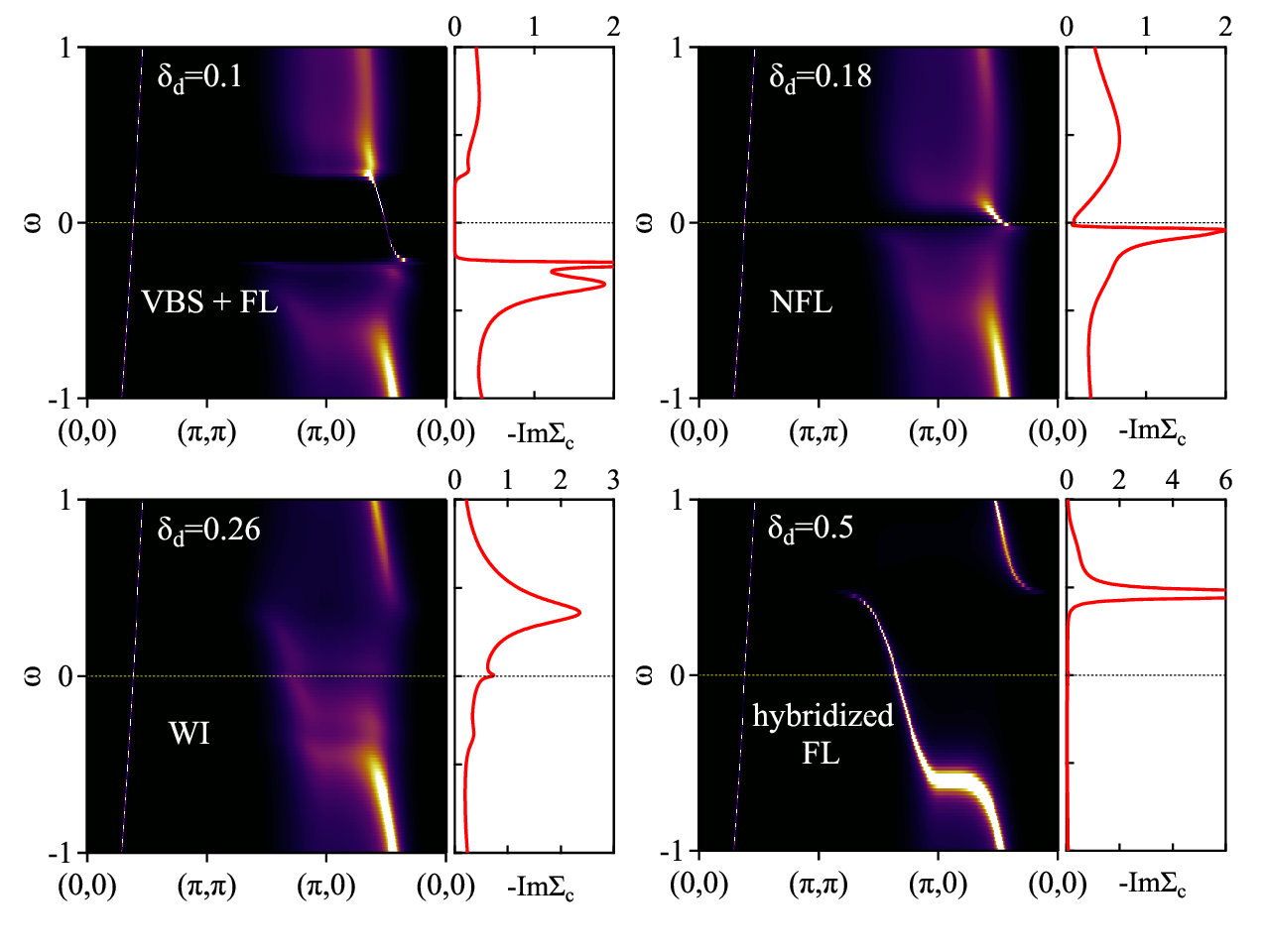}
	\caption{Evolution of the $d_{x^2-y^2}$ spectral function and the imaginary part of self-energy $-\text{Im}\Sigma_c(\omega)$ with $d_{z^2}$ hole doping $\delta_d$ for $J=V=0.5$ and $U=7$. Only nearest neighbor hopping is used to obtain the simplified $d_{x^2-y^2}$ spectra. The four doping levels correspond to VBS + FL ($\delta=0.1$), NFL ($\delta=0.18$), WI ($\delta=0.26$) and hybridized FL ($\delta=0.5$) normal states, respectively.  }
	\label{fig:Ac}
\end{figure}

\begin{figure}[b]
	\centering\includegraphics[scale=0.55]{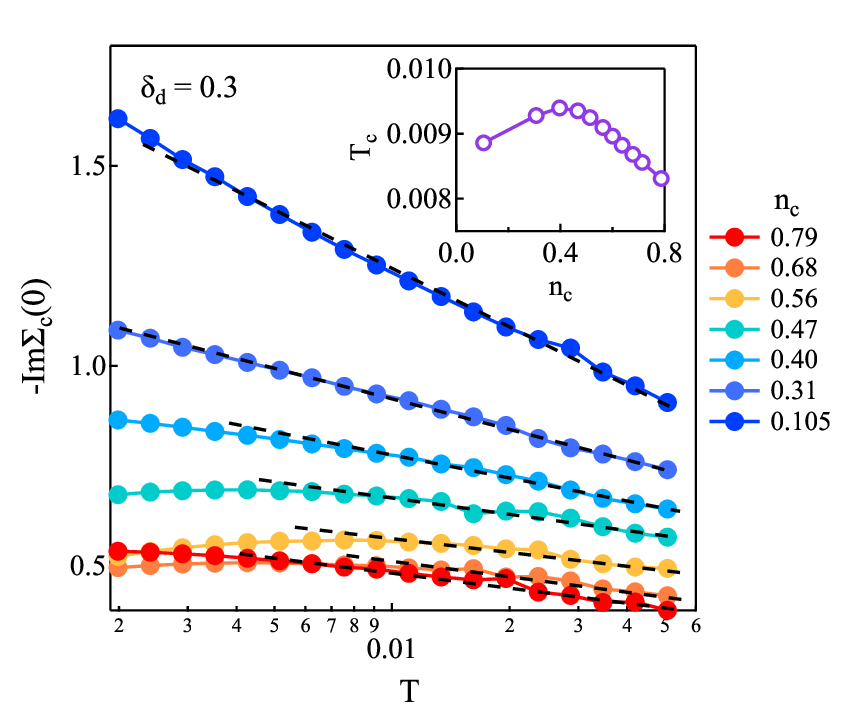}
	\caption{Evolution of the weakly insulating behavior reflected in the $d_{x^2-y^2}$ self-energy, $-\text{Im}\Sigma_c(0)$, with respect to the $d_{x^2-y^2}$ occupation number $n_c$ at a fixed $d_{z^2}$ hole doping $\delta_d=0.3$. The dashed lines mark the $-\ln T$ behavior. The inset shows $T_c$ as a function of $n_c$ at $\delta_d=0.3$. Other parameters are $J=V=0.5$ and $U=7$. }
	\label{fig:WI}
\end{figure}

\subsection{III. Normal state evolution with $d_{z^2}$ hole doping and $d_{x^2-y^2}$ occupation}
The nature of the different normal states is reflected in the $d_{x^2-y^2}$ spectra and self-energies. Figure \ref{fig:Ac} plots the results using the simplified bare dispersion of $d_{x^2-y^2}$ electrons for four representative $d_{z^2}$ hole doping levels ($\delta_d=0.1, 0.18,  0.26, 0.5$) at $J=V=0.5$, $U=7$. For all cases, we observe a sharp spectrum along the zone diagonal direction ($(0,0)\rightarrow (\pi,\pi)$) but highly renormalized spectra along $k_x$ ($(\pi,0)\rightarrow (0,0)$) due to the special anisotropic hybridization. For $\delta_d=0.1$, one sees clearly a sharp $d_{x^2-y^2}$ spectrum within a finite gap of $-\text{Im}\Sigma_c(\omega)$ around the Fermi energy, indicating nearly decoupled $d_{x^2-y^2}$ electrons from the $d_{z^2}$ VBS at low temperature. The self-energy gap diminishes into a dip at $\delta_d=0.18$, where the quasiparticle spectrum along $k_x$ is quickly smeared and turns into a continuum away from the Fermi energy, indicating strong scattering between $d_{z^2}$ and $d_{x^2-y^2}$ electrons. The strong inter-orbital coupling is responsible for the NFL behavior around optimal $T_c$. For $\delta_d=0.26$, the gap (dip) disappears completely and the quasiparticle spectrum along $k_x$ is strongly smeared due to the large self-energy, causing the $-\ln T$ weakly insulating behavior associated with incoherent Kondo scattering. Upon further doping, a well-defined quasiparticle band appears at large $\delta_d=0.5$, showing clear hybridization features above the Fermi level, which is identified as a hybridized Fermi liquid. The Fermi surface topology changes dramatically as the system evolves from the NFL to the hybridized FL state across the WI region, induced by the gradual delocalization of $d_{z^2}$ interlayer valence bonds and the formation of hybridized quasiparticle bands upon increasing doping. Similar evolution also appears for $d_{z^2}$ electron doping.  

Such normal state evolution is insensitive to the $d_{x^2-y^2}$ occupation as well as the detailed form of its bare dispersion. Figure \ref{fig:WI} plots the temperature dependence of $d_{x^2-y^2}$ self-energy, $-\text{Im}\Sigma_c(0)$, with varying $d_{x^2-y^2}$ occupation $n_c$ at a fixed $\delta_d=0.3$. Within a wide range of $n_c\in [0.1, 0.8]$, the $-\ln T$ weakly insulating behavior can always be identified with only slight variation in the temperature range. We also show in the inset the calculated $T_c$ as a function of $n_c$ for the same parameters ($J=V=0.5$, $U=7$, $\delta_d=0.3$), and find that the maximal $T_c$ occurs roughly around quarter filling ($n_c\approx 0.4-0.5$).